# A Distributed Ledger-Enabled Interworking Model for the Wireless Air Interface

Steven Platt
Department of Information and Communication
Technologies Engineering
Pompeu Fabra University
Barcelona, Spain
steven.platt@upf.edu

Miquel Oliver
Department of Information and Communication
Technologies Engineering
Pompeu Fabra University
Barcelona, Spain
miquel.oliver@upf.edu

*Abstract*— While direct allocation of spectrum and evolved medium access protocols provide a base for ubiquitous wireless connectivity, the existing TCP/IP and OSI models were designed for wired networks and do not address open interconnection of air interfaces. Without an interconnection model for the air interface, existing network designs continue to tie wireless medium access to that of the backhaul provider for ownership of access and identity trust, resulting in limitations on functionality and coverage.

In this paper, we propose a novel solution to access ownership and identity trust by extending the TCP network standard, under a new model we propose, named TCP-Air which integrates distributed ledger technologies directly at the air interface. Further, we present two use cases of the TCP-Air model, demonstrating applications not feasible under existing permissioned-access network designs.

*Keywords*— *Access control, blockchain, distributed ledger, Internet of Things (IoT), cognitive radio, vehicle networks*

I. INTRODUCTION

Today, wireless networks continue to be treated as permissioned gateways of access to a wired backhaul network, rather than an independent environment - leading to gaps in coverage and access. To address these shortcomings, TCP-extending protocols such as Ad Hoc On-Demand Distance Vector (AODV), and Optimized Link State Routing (OLSR) were introduced as part of the mobile ad-hoc networks paradigm (MANETS). These designs have since failed to gain market adoption [1], [2], in part due to a failure to address the structure of permissioning network access, required ahead of ad-hoc routing.

In this paper, we propose a novel interworking model for the wireless air interface, built upon decentralized blockchain distribution, named "TCP-Air". The proposed framework splits interworking functions of the air interface into four model layers, able to operate wholly independent of the wired network functions underneath. The layers, as defined are: Application, Identity, and Spectrum, atop a base layer defined as Ledger - effectively substituting transport functions which bind the air interface and backhaul today.

The main contributions of the proposed model are:

*Mobility:* TCP-Air allows full abstraction of air interface functions from those of a wired backhaul, enabling mobility independent of physical layers underneath.

*Security:* Using Identity-Based Networking, and immutable ledger distribution, TCP-Air enables higher security, associated to a device identity carried between networks.

*Autonomy:* Combining the additional context of device behaviour, TCP-Air enables autonomous access of physical devices, resolving limitations of existing permissioned network access.

The remaining data of this paper is organized in the following sections. Section II provides information of the state of the art, providing context of recent research into Identity-Based Networking, and Distributed Ledger technologies which informed the TCP-Air design. Section III defines the proposed TCP-Air interworking model and details its four component layers: application, identity, spectrum, and ledger. Section IV details TCP-Air in generalized application, detailing two use cases designed under the model. In Section V we discuss the limitations and implications of the proposed framework, and in Section VI we provide a summary of the papers contributions and planned direction for further research.

II. STATE OF THE ART

*A. Identity-Based Networking*

Under TCP/IP, the IP address serves dual purpose, as both the machine identity, and a basis for routing to a device in a network. While designed for universality, the IP address introduces security vulnerabilities, because the address itself is not backed by anything verifiable and can be easily spoofed.

In 2015, the IETF adopted draft RFC 7401 [3] for the creation of the Host Identity protocol (HIP). This protocol intends to insert an identity mechanism between the network and transport layers of TCP/IP, effectively isolating the two functions of the IP address, while retaining backwards and forwards compatibility. HIP achieves this by replacing existing 'IP Address + Port' routing used at the higher layers of

TCP/IP, with a new 'Host Identity + Port' pairing (fig. 1) generated using a Diffie-Hellman method of public key exchange [3]. Under Diffie-Hellman, two communicating parties exchange cryptographic public keys in order to form a shared private key without requiring a prior trust. Because this new identity is verifiable, a network implementing HIP is significantly more secure against man-in-the-middle and DDoS style attacks, which exploit the unverifiability of a traditional IP address [3]. Placing identity as the centre of network design in this manner often referred to as "Identity-Based Networking".

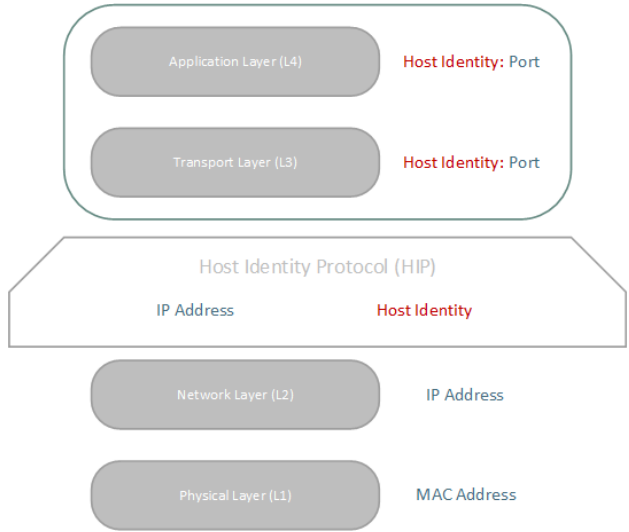

Figure. 1. The modified Host Identity and port pairing applied with Host Identity Protocol.

### B. Distributed Ledger and Internet 3.0

The original Bitcoin whitepaper was published in 2008, and detailed the design of a digital currency system that removed the need of a trusted third party for the verification of transactions [4]. The system did this using a peer-to-peer distribution of a universal ledger. Bitcoin's ledger is designed as a never ending chain structure, where new data being added to the chain requires combining the timestamp of the last transaction, along with a hash of the new data being appended to the ledger (fig 2). The resulting aggregate hash gives the ledger its chain structure, and is seen as immutable and highly secure, as a recomputation of all subsequent work is required in order to modify old transactions on the chain [5], [6], [7]. The later ubiquity of Bitcoin helped popularized the terms Blockchain and Distributed Ledger.

While Bitcoin itself was designed as a digital currency, the blockchain structure of its underpinning ledger has been applied in a number of applications requiring high trust without direct ownership. This class of applications are increasingly termed as "Internet 3" indicating a belief that the next generation of internet development will exist in this decentralized model. Filecoin [8] is an example of such Internet 3 class application. Filecoin is a decentralized network storage platform. Built on top of a distributed ledger; the Filecoin system allows users to pay for storage of data on the network or earn payment through hosting the files of others. Files stored in the Filecoin network use data sharding and peer-to-peer distribution to ensuring file contents cannot be read by network participants not directly owning the file.

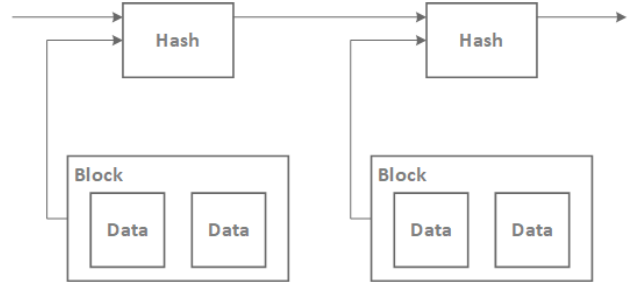

Figure. 2. Bitcoin blockchain structure

### III. THE TCP-AIR INTERWORKING MODEL

TCP-Air is an abstract model designed to provide a framework for direct interworking of the wireless air interface by combining existing TCP/IP and wireless medium access functions, with new services provided through host identity and distributed ledger technologies.

Similar to the TCP/IP and OSI model, TCP-Air is modelled using a layered architecture [9], with each interconnecting system being composed of subsystems, where equivalent subsystems exist in the same layer of the model and interactions occur only between subsystems at adjacent layers.

In total, the TCP-Air model is consisting of four layers, they are: Application, Identity, Spectrum, and Ledger.

TCP-Air is not designed as a replacement for the existing TCP/IP standard, rather as a parallel model, in a relationship similar to that of TCP/IP and the OSI model, while being purpose built to handle interworking of air interface networks.

The following sections provide detail on the function of each layer of the TCP-Air model (table I), split into two sections: Layer Function, and Services Provided.

### A. Definitions

- Air Interface: The radio managing interconnection between access points and other physical devices, which use air as the transport medium.

- Peer: Two air interfaces which have the ability to bi-directional exchange data using air as the transport medium.

- Supplication: The registration of an access point as client to another access point.

- Block: A single unit of storage for data written to a ledger.
- Chain: The total of block data, which has been cryptographically linked. (a chain of blocks).
- Ledger: The aggregate of block storage, chained and distributed throughout a network. (distributed ledger).
- Route: The stored path to an air interface peer within separately managed networks that is connected through federation.

TABLE I. TCP-AIR MODEL LAYERS

| Layer Number | TCP-Air Model Layers | |
|---|---|---|
| | *Layer Name* | *Layer Functions\** |
| 4 | Application | Internet Protocol, User Interface |
| 3 | Identity | Host Identity, Profiling, Access and Authorization |
| 2 | Spectrum | Spectrum Addressing, Wireless Medium Access, Spectrum Sensing, Peer Supplication, Ledger Termination, TCP/IP Termination |
| 1 | Ledger | Block Store, Route Store, Block Distribution, Block Cache, Chain Connection, Chain Management, Accounting |

*Table is not inclusive of all possible model layer functions

### B. Application Layer

*1) Layer Function:* The application layer handles peer-to-peer and client-server communications for applications and services using Internet Protocol (IP), as well as surfacing this data for user interaction.

*2) Services Provided:*

*a) Internet Protocol:* Handling of communications between peer protocols functioning in the Internet Protocol suite, including, but not limited to: HTTP, FTP, DNS, SSL, IMAP, NTP, SIP, and SMTP.

*b) User Interface*: Handling of data presentation, and manipulation for services allowing end-user exposure or interaction.

### C. Identity Layer

*1) Layer Function:* Execution of identity and authentication functions within the TCP-Air model.

*2) Services Provided:*

*a) Host Identity:* Assignment of host identity for routing and termination of application, network, and chain connections. Correlation of cryptographic identity, hardware address, and/or IP address.

*b) Profiling:* Correlation of traffic patterns, access patterns, mobility patterns, and device physical attributes.

*c) Access and Authorization:* The identity layer is responsible for granting and revoking access to network, spectrum, and chain resources.

### D. Spectrum Layer

*1) Layer Function:* The spectrum layer handles all functions to enable connection within wireless spectrum.

*2) Services Provided:*

*a) Spectrum Addressing:* Legacy MAC and IP addressing of physical devices accessing a given air interface.

*b) Wireless Medium Access:* Wireless channel assignment, power, flow, and transmission controls, beam-forming, contention resolution, and quality of service.

*c) Spectrum Sensing:* Scanning of compatible air channels and surrounding devices [10].

*d) Peer Supplication:* Registering as supplicant to unmanaged air interface peers. Routing and and peering of data outside a given air interface.

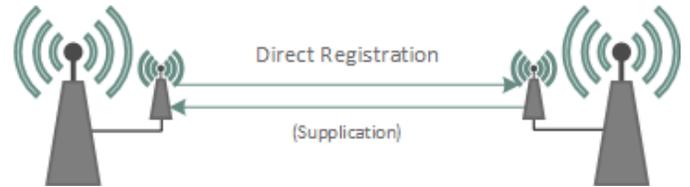

Fig. 3. Air interface peering through direct supplication

*e) Ledger Termination:* The spectrum layer handles routing and delivery of data and connection to the ledger layer for immutable storage [11].

*f) TCP/IP Termination:* Routing and termination of data and connection to the broader internet.

### E. Ledger Layer

*1) Layer Function:* The ledger layer is the communications path of the TCP-Air model and provides abstraction from the routing and topology of a wired backhaul. The ledger layer enables seemless mobility, and higher security [6], through the syncing, caching, and distribution of network permissions among participating air interface termination points [11], [12].

*2) Services Provided:*

*a) Block Store:* Assignment, hashing, and storage of data onto chains.

*b) Route Store:* Recording of air interfaces, network routes, peer air interface routes, and/or cryptographic identity data for physical devices transacting on a given chain.

*c) Block Distribution:* Peer-to-Peer propagation of block data among air interface termination points, transacting on a given chain.

  *d) Block Cache:* Temporary storage of chain data with highest probability of near-term access.

  *e) Chain Connection:* Connectivity between physical endpoints accessing the same service whose permissions are distributed on the ledger layer.

  *f) Chain Management:* Management of network participation. Orchestration of chain functions, including contracts.

  *g) Accounting:* Immutable storage of physical device access and behavior profile data.

## IV. Sample Use Case: Vehicle Networks

Despite the maturity of ITS [13], [14] and vehicle network research [15], there has not been developed consensus for handling permissioned network access in a manner that enables pervasive adoption. For this reason, a vehicle network example is chosen to demonstrate the resolution of coverage limitations through adoption of a TCP-Air model design.

### A. Network Design

The design assumes random distribution of roadside units on a road network spanning multiple municipalities. Each municipality operates independently, without a trust to authorize vehicles arriving from outside of its zone. Figure 4 shows the interaction among model layer functions.

### B. Layer Functions

*1) Application Layer:* To demonstrate expanded coverage, the application in this example is the general internet. No restriction is placed on routing in this design.

*2) Identity Layer:* Access to the network is granted autonomously. To establish identity in the trustless environment, a host identity is created for each vehicle. This identity serves both as an aggregation point for behavior data, and as an abstraction for routing as vehicle IP addressing changes in network handoff between municipalities. With an identity abstracted from network ownership, a profile is created to learn devices with mobility patterns matching that of the road network - determined by patterns roadside unit adjacency, distribution and variation of speed. A final filter of hardware address (MAC) is applied to remove devices which are not produced by known vehicle manufacturers. The resulting host identity profile is set on an expiry, to allow pruning of retired vehicles and restrict anomalous vehicles, not matching established identity histories and patterns of behavior.

*3) Spectrum Layer:* Utilizing spectrum scanning [10], the network reports device adjacencies. Through RSSI-based network localization [16][17], vehicle speed and direction are deduced. As the TCP termination point, and the layer handling medium access, an open SSID is broadcast with its default VLAN utilizing a null route. Vehicles learned and authorized by network behavior profiling are automatically placed into a routed VLAN to the broader internet.

*4) Ledger Layer:* Rules governing access to the vehicle network are stored as a smart contract on a single ledger chain [18]. The same chain stores vehicle behavior, host identity, and access authorizations. Data stored on the chain is immutable and propagated among participating roadside units in the multi-municipality network. Additional chains may be added to the ledger, for future services, such as safety and hazard notifications, without managed network access or identity trust [11], [12].

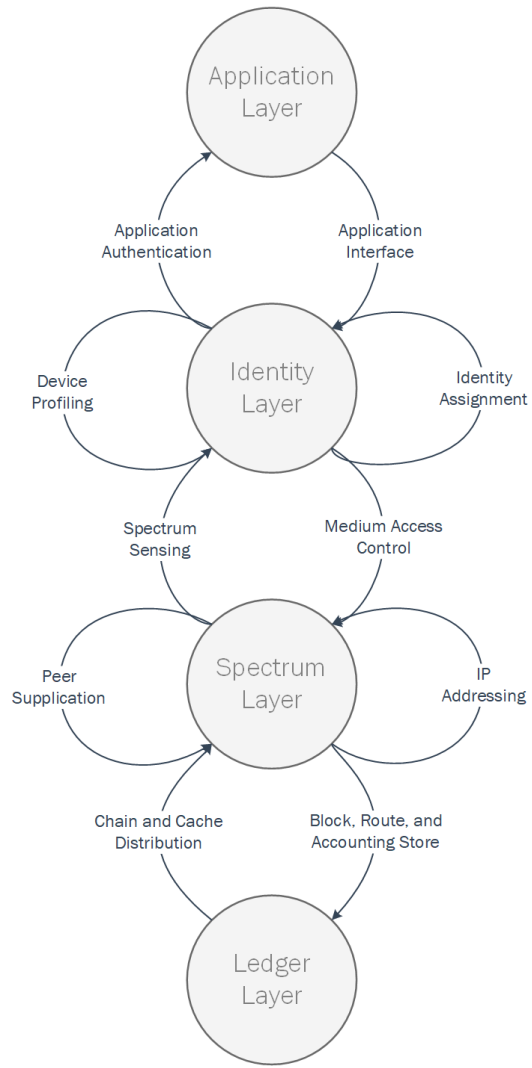

Fig. 4.: Layer interaction model of TCP-Air

### C. Further Implementation

Through reuse of layer functions presented in our vehicle network example, the extended functionality of wireless emergency location services is possible. This additional use case is presented as a high-level proposal, to add further context for TCP-Air use as a generalized model.

*1) Wireless Emergency Location Services:* Under government regulation, regions such as Europe and the United States, required the inclusion of accurate location data with calls placed to emergency services from wireless networks [19]. In cellular networks, this is implemented using network localization among neighboring towers. A limitation of this implementation is introduced with elevation within buildings, due to refraction and reflection experienced among signals used to calculate location. For extend network functionality, to include accurate elevation, we propose use of access federation functions within the TCP-Air model among 802.11 access points within the building environment. In this design, further localization among 802.11 access points allow deduction of floor level, through the awareness and sorting of access point signal adjacency at a given building address.

## V. Discussion

Implementation of functions described in the model are based on several assumptions.

### A. Equal network participation and traffic distribution

When connecting unmanaged networks, there arises a possibility of asymmetry within network traffic, for example: disproportionality due to network size, device numbers, or physical device distribution. In our example of federated coverage among municipalities, it is possible that a municipality hosting an urban hub has a network density exceeding the capacity of peer networks. It is possible to program balance into such traffic distribution, through routing modification, caching of interactions, quality of services measures, or smart contract functions programmed within the ledger. Such mechanisms of system-wide resource management require further research.

### B. Scalability of the distributed ledger

There are many variations of ledge implementation, each with unique benefits and dependencies. The model as presented does not make an explicit choice of ledger technology, rather it amalgamates existing ledger technologies to prevent association or inheritance of characteristics existing in any single ledger. This limits specificity within the example designs, but open an area of further research and resolution of which performance measures of distributed ledgers best suite a network infrastructure environment [20].

### C. No association between physical device profile and user identity

Examples provided for TCP-Air implementation, enable autonomous access built on identity profiling. Because the system uses spectrum scanning of the ambient environment, there is no implicit ability to opt in or out of detection. Further, the profile data is collected over time and made immutable through its storage within a distributed ledger. Although profile data is collected and correlated based on physical device behaviours, the extent to which these devices can be associated to unique users may invoke additional privacy restriction on such model functions. As a base, the model assumes no association between physical device and user identity.

## VI. Conclusion and Future Work

In this paper we identify the structure of permissioned network access and trust of identity, as characteristics which have prevented direct interworking of the wireless air interface in current designs. To address these limitation, we propose a new interworking model, named TCP-Air, combining existing function of wireless networks, with two new interworking layers handling identity and immutable storage on a distributed ledger. Additionally, the paper outlines two example deployment under the model, which enables autonomous network access and fluid mobility to create a pervasive vehicle network infrastructure, as well as enhanced functionality for wireless emergency location services. For future work, we are interested to investigate specific implementations of the ledger base layer protocols, to examining performance and scalability within a live environment implementation.